\documentclass[a4paper,12pt, epsfig]{article}
\def\letter{0}\def\pr{0}
\pdfoutput=1 
\usepackage{epsfig}
\usepackage{epstopdf}
\usepackage{graphicx}
\usepackage{ifthen}

\usepackage{amsmath}
\usepackage[psamsfonts]{amssymb}
\usepackage{euscript}

\usepackage{latexsym}
\usepackage[arrow,matrix,curve]{xy}

\newskip\humongous \humongous=0pt plus 1000pt minus 1000pt

\newif\ifdtup

\def\,{\hspace{-.1cm}}
\def\hsp{,\hspace{.7cm}}

\def\fc#1#2 {\frac{n}{q}#1\frac{n}{q}#2}

\def\hf{H\p_{f_0}}

\newcommand{\vac}{\ensuremath{|0\rangle}}
\newcommand{\stt}{\ensuremath{|\psi\rangle}}

\renewcommand{\sinh}{\textrm{sinh}}
\renewcommand{\cosh}{\textrm{cosh}}
\renewcommand{\tanh}{\textrm{tanh}}
\newcommand{\sech}{\textrm{sech}}
\newcommand{\csch}{\textrm{csch}}

\def\exp#1{\hbox{\rm exp}\left(#1\right)}

\renewcommand{\theequation}{\arabic{section}.\arabic{equation}}
\renewcommand{\(}{\begin{equation}}
\renewcommand{\)}{end{equation} \vspace{-.05in}\linebreak}

\newcounter{saveeqn}
\newcounter{savealpheqn}

\newcommand{\alpheqn}{\setcounter{saveeqn}{\value{equation}}%
  \stepcounter{saveeqn}\setcounter{equation}{0}%
  \renewcommand{\theequation}{\mbox{\arabic{section}.\arabic{saveeqn}
\alph{equation}}}
  \renewcommand{\)}{\end{equation}}}
\def\part#1{\frac{\partial}{\partial{#1}}}%
\def\group#1{\refstepcounter{equation}\setcounter{saveeqn}
 {\value{equation}}%
  \label{#1}\setcounter{equation}{0}%
\renewcommand{\theequation}{\mbox{\arabic{section}.\arabic{saveeqn}
\alph{equation}}}
  \renewcommand{\)}{\end{equation}}}
\newcommand{\reseteqn}{\setcounter{equation}{\value{saveeqn}}%
  \renewcommand{\theequation}{\arabic{section}.\arabic{equation}}%
  \renewcommand{\)}{\end{equation}}}

\newcommand{\aalpheqn}{\setcounter{saveeqn}{\value{equation}}%
  \stepcounter{saveeqn}\setcounter{equation}{0}%
  \renewcommand{\theequation}{\mbox{
        \Alph{subsection}.\arabic{saveeqn}\alph{equation}}}
   \renewcommand{\)}{\end{equation}}}
\newcommand{\areseteqn}{\setcounter{equation}{\value{saveeqn}}%
  \renewcommand{\theequation}{\Alph{subsection}.\arabic{equation}}%
  \renewcommand{\)}{\end{equation}}}

\renewcommand{\thefootnote}{\alph{footnote}}
\renewcommand{\(}{\begin{equation}}
\renewcommand{\)}{\end{equation}}
\newcommand{\ba}{\begin{eqnarray}}
\newcommand{\ea}{\end{eqnarray}}

\renewcommand{\b}{\beta}

\newcommand{\cbp}{\mathop{\vtop{\ialign{##\crcr
   $\hfil\displaystyle{}\hfil$\crcr\noalign{\kern-13pt\nointerlineskip}
   \BIG{)}\hskip0pt\crcr\noalign{\kern3pt}}}}}
\newcommand{\pa}{\mathop{\vtop{\ialign{##\crcr

$\hfil\displaystyle{\oplus}\hfil$\crcr\noalign{\kern+1pt\nointerlineskip
}
   \hspace{.08in}$^{\alpha=0}$\hskip6pt\crcr\noalign{\kern3pt}}}}}
\renewcommand{\hsp}{,\hspace{.3in}}
\newcommand{\p}{^\prime}

\def\D{\ensuremath{{\cal D}}}

\catcode`\@=11
\def\vereq#1#2{\lower3pt\vbox{\baselineskip1.5pt \lineskip1.5pt
\ialign{$\m@th#1\hfill##\hfil$\crcr#2\crcr\sim\crcr}}}
\catcode`\@=12

\renewcommand{\(}{\begin{equation}}
\renewcommand{\)}{\end{equation}}

\def\pin#1{\int \frac{d#1}{2\pi}}
\def\ppin#1{\int\hspace{-17pt}\sum \frac{d#1}{2\pi}}
\def\ppink#1{\int\hspace{-17pt}\sum\frac{d^{#1}k}{(2\pi)^{#1}}}
\def\dint{\int\hspace{-12pt}\sum }
\def\pink#1{\int \frac{d^{#1}k}{(2\pi)^{#1}}}

\def\Bd#1{B^\dag_{k_{#1}}}

\def\df{\mathcal{D}_{f}}
\def\dF{\mathcal{D}_F}

\def\I{\mathcal{I}}

\def\os{\omega_S}

\def\as{|\alpha;\sigma\rangle}
\def\asb{\langle\alpha;\sigma|}

\newcommand{\beas}{\begin{eqnarray*}}
\newcommand{\eeas}{\end{eqnarray*}}

\newcommand{\bquo}{\begin{quote}}
\newcommand{\enqu}{\end{quote}}

\newcommand{\g}{{\mathfrak g}}

\def\ch{{\mathcal{H}}}

\def\ok#1{\omega_{k_{#1}}}

\def\V#1{V^{(#1)}(gf(x))}

\newcommand{\beq}{\begin{equation}}
\newcommand{\eeq}{\end{equation}}
\newcommand{\bea}{\begin{eqnarray}}
\newcommand{\eea}{\end{eqnarray}}

\newskip\humongous \humongous=0pt plus 1000pt minus 1000pt

\newif\ifdtup

\catcode`\@=11

\ifthenelse{\equal{\letter}{0}}{ 

\@addtoreset{equation}{section}
\def\theequation{\arabic{section}.\arabic{equation}}

\def\@normalsize{\@setsize\normalsize{15pt}\xiipt\@xiipt
\abovedisplayskip 14pt plus3pt minus3pt%
\belowdisplayskip \abovedisplayskip
\abovedisplayshortskip \z@ plus3pt%
\belowdisplayshortskip 7pt plus3.5pt minus0pt}

\def\small{\@setsize\small{13.6pt}\xipt\@xipt
\abovedisplayskip 13pt plus3pt minus3pt%
\belowdisplayskip \abovedisplayskip
\abovedisplayshortskip \z@ plus3pt%
\belowdisplayshortskip 7pt plus3.5pt minus0pt
\def\@listi{\parsep 4.5pt plus 2pt minus 1pt
      \itemsep \parsep
      \topsep 9pt plus 3pt minus 3pt}}

\relax

\def\section{\@startsection{section}{1}{\z@}{3.5ex plus 1ex minus  .2ex}{2.3ex plus .2ex}{\large\bf}}

\def\thesection{\arabic{section}}
\def\thesubsection{\arabic{section}.\arabic{subsection}}

\def\appendix{\setcounter{section}{0}
 \def\thesection{Appendix \Alph{section}}
 \def\thesubsection{\Alph{section}.\arabic{subsection}}
 \def\theequation{\Alph{section}.\arabic{equation}}}
\renewcommand{\theequation}{\arabic{section}.\arabic{equation}}

}{
\renewcommand{\theequation}{\arabic{equation}}

} 

\begin{document}
\def\thefootnote{\fnsymbol{footnote}}
\def\thetitle{Moving Kinks and Their Wave Packets}
\def\autone{Jarah Evslin}
\def\auttwo{Hengyuan Guo}
\def\affa{Institute of Modern Physics, NanChangLu 509, Lanzhou 730000, China}
\def\affb{University of the Chinese Academy of Sciences, YuQuanLu 19A, Beijing 100049, China}

\title{Titolo}

\ifthenelse{\equal{\pr}{1}}{
\title{\thetitle}
\author{\autone}
\affiliation {\affa}
\affiliation {\affb}

}{

\begin{center}
{\large {\bf \thetitle}}

\bigskip

\bigskip


{\large \noindent  \autone{${}^{1,2}$} }


\vskip.7cm

1) \affa\\
2) \affb\\

\end{center}

}

\begin{abstract}
\noindent
Recently a linearized perturbation theory has been formulated for soliton sectors of quantum field theories.  While it is more economical than alternative formalisms, such as collective coordinates, it is currently limited to solitons which stay close to a base point, about which the theory is linearized.  As a result, so far this formalism has only been applied to stationary solitons.  In spite of this limitation, we construct kink states with fixed nonzero momenta and also moving, normalizable kink wave packets.  The former are nonnormalizable, coherent superpositions of kinks at all spatial positions and are simultaneous eigenstates of the Hamiltonian and the momentum operator.  The latter are localized about a single, moving classical solution.  To understand the wave packets, we calculate several simple matrix elements.


\end{abstract}

%
\setcounter{footnote}{0}
\renewcommand{\thefootnote}{\arabic{footnote}}

\ifthenelse{\equal{\pr}{1}}
{
\maketitle
}{}

\section{Introduction}

\subsection{Motivation}

Linearized soliton perturbation theory \cite{me2loop} allows the efficient\footnote{The leading quantum corrections can be computed efficiently in great generality using spectral methods, recently reviewed in Ref.~\cite{weigrev}.} calculation of states \cite{mestato}, masses \cite{memassa} and instantaneous accelerations \cite{chris} of solitons in nontrivial backgrounds.   However so far it has one major limitation:  The solitons cannot move.  As a result, the trajectory of a soliton in a nontrivial background cannot be found, as once it begins to move the corresponding state is no longer known.  Also form factors cannot be calculated, as these involve states with nonvanishing momentum.  Finally, in models without Poincar\'e invariance, such as those with impurities \cite{impure19}, it has not yet been possible to include quantum corrections into moduli space truncated Hamiltonians \cite{muri,moduli} because the energy dependence on the soliton velocity is not known.

This limitation may seem inevitable, as the method begins with a unitary transformation of the Hilbert space which is determined by the choice of a single point in the soliton's moduli space.  In this note we provide two distinct solutions to this problem.   More precisely, we present two constructions of states corresponding to solitons with nonzero momentum.  The first construction is simply a boost of the construction of a stationary soliton.  Although the boosted soliton has momentum, it is a momentum eigenstate and so is translation invariant up to a phase.  This implies that the kink state includes a uniform superposition of kink positions over the entire space. Therefore it does not move, and there is no contradiction with the above intuition.  The second construction uses a normalizable wave packet of solitons localized about some point in moduli space.  This  is not an exact eigenstate of the momentum nor of the Hamiltonian, and so it does move.

These two constructions correspond to two distinct physical configurations, both of which are realized in Nature.  In QCD, in the large $N$ approximation, baryons are described by Skyrmions \cite{skyrme,wittenskyrme,smorg}.  Baryon scattering is described by the scattering of solitons in wave packets which are nearly momentum eigenstates, and so are well described by plane waves.  In particular, their wave packet size is much large than their Fermi-scale radius.   This corresponds to our first construction.  On the other hand, often a soliton position is constrained to greater precision than the soliton size itself.  Such semiclassical solitons have a quantum profile that resembles the corresponding classical field theory solution.  This second case includes solitonic dark matter \cite{medark,lorodark} as well as many examples in condensed matter physics, beginning historically with Abrikosov vortices \cite{abrik} on an observed lattice and also many solitons in quantum optics, such as \cite{optix}.  

\subsection{Background}

A quantum theory is defined by a Hamiltonian operator $H$ and a Hilbert space on which it acts.  The stationary states are eigenvectors of $H$.  Let us consider a Schrodinger picture quantum field theory of a single scalar field $\phi(x)$, where $x$ is a point in space.  In this case, the operators $\phi(x)$ at each $x$ and their conjugate momenta $\pi(x)$ are a basis of the space of all operators in the theory.  In particular, the Hamiltonian is constructed from these operators.  

In the quantum field theory, the operators satisfy the canonical commutation relations $[\phi(x),\pi(x)]=i\hbar\delta(x)$.  We will generally set $\hbar=1$.  However, setting $\hbar=0$ one arrives at the corresponding classical field theory.  If the classical equations of motion derived from this Hamiltonian have a nontrivial, stable, stationary solution $\phi(x,t)=f(x)$, then one may ask what state $|K\rangle$ in the quantum theory corresponds to this classical configuration.  More generally, one may consider small perturbations about this classical solution and wonder to which quantum states they correspond.  We will refer to such states as the $f(x)$-sector.  

Old fashioned perturbation theory expands the field $\phi(x)$ about zero and so does not yield states in the $f(x)$-sector if $f(x)$ is not identically zero.  Therefore the usual approach \cite{dhn2} to studying the $f(x)$-sector is to decompose the field into a classical part and a quantum part $\phi(x)-f(x)$, rewrite the defining Hamiltonian as a kink Hamiltonian for this quantum part and try to diagonalize the kink Hamiltonian.  The potential problem with this approach is that quantum field theories generally have divergences that require regularization, and simple regularization schemes such as an energy cutoff do not commute with the transition from the defining to the kink Hamiltonian \cite{rebhan}.

Recently this problem has been solved in Ref.~\cite{mekink} in a rederivation of the manifestly finite kink Hamiltonian of Ref.~\cite{cahill76}.  The regularized defining Hamiltonian $H$ defines the theory, and so the regularized kink Hamiltonian $H\p$ is defined to be similar, in fact unitarily equivalent, to the regularized defining Hamiltonian.  This guarantees that they will have the same spectrum, and so one may first perturbatively solve the $H\p$ eigenvalue problem and then use the unitary map to create $H$ eigenvectors from $H\p$ eigenvectors.

Concretely, one defines the unitary displacement operator
\beq
\df={\rm{exp}}\left(-i\int dx f(x)\pi(x)\right) \label{df}
\eeq
which commutes with $\pi(x)$ but shifts $\phi(x)$
\beq
\phi(x)\df=\df\left(\phi(x)+f(x)\right).
\eeq
Then the kink Hamiltonian $H\p$ and even the kink momentum $P\p$ are defined by
\beq
H\p=\df^\dag H\df\hsp
P\p=\df^\dag P\df
 \label{hpd}
\eeq
where $P$ is the momentum operator.  Intuitively, this unitary equivalence reexpresses the operators in terms of the quantum field $\phi(x)-f(x)$ as in the traditional approach, but unlike the traditional approach it never changes the spectrum as $H\p$ and $H$ are related by a similarity transformation (\ref{hpd}).  We remind the reader that $H$ is already regularized, and so $H\p$ will be automatically regularized.

The strategy then is to use perturbation theory to obtain the desired eigenstate $|\psi\rangle$ of $H\p$ and then to act on it with $\df$ to obtain to corresponding eigenstate $\df|\psi\rangle$ of $H$.  In other words, one first performs $\df^\dag$ on the original Hilbert space yielding the kink Hilbert space.  Next one diagonalizes the kink Hamiltonian perturbatively in the kink Hilbert space.  Finally one performs $\df$ to return to the original, defining Hilbert space. 

This application of perturbation theory is somewhat complicated in a Poincar\'e-invariant theory because translation invariance leads to an infinity of soliton solutions, and therefore a gapless spectrum, leading to the usual infrared divergences in the perturbative expansion.  These divergences are usually eliminated using the collective coordinate approach \cite{gjscc}, which consists of a nonlinear canonical transformation which disentangles the problematic zero-mode.  

Recently, a much more economical approach has been proposed \cite{me2loop} in which one instead first solves the $P\p$ eigenvalue equation in perturbation theory.  Once this is done, the problematic degeneracy is removed and one then imposes the $H\p$ eigenvalue equation.  This avoids nonlinear transformations and in fact simplifies the problem, as $P\p$ is simpler than $H\p$ and its form is independent of the interactions.  

However, the price of solving the $P\p$ eigenvalue equation only perturbatively is that one is effectively expanding about a base point in the moduli space, and so the series found does not converge, even in the sense of an asymptotic series, far from this base point.  To be able to construct states near that base point one may conclude that the kink cannot move, and so all previous studies of this formalism have restricted attention to stationary kinks.

\subsection{Outline}

In Sec.~\ref{boostsez}, we will find that one can nonetheless construct a kink state with nonvanishing momentum, an eigenvector of the momentum operator.   This is reasonable as such kink plane-waves are, up to a phase, time-independent.  This is because although they have nonzero velocity, they are everywhere, and so they do not move.   

This is potentially useful for calculating energy spectra but still not sufficient for problems such as scattering, for which one wants a localized soliton corresponding to a normalizable state with finite matrix elements.  Such localized, normalizable states have not yet been constructed even for solitons with vanishing momentum.  In Sec.~\ref{pacsez} we construct such normalizable kink wave packets.  They indeed do move, and so they are not exact Hamiltonian eigenstates, which are necessarily time-independent.  However, as they are normalizable, they allow us to compute matrix elements for the first time using linearized perturbation theory.

\section{The Kink Hamiltonian Eigenvalue Problem} \label{revsez}

In this section we review the solution of the eigenvalue problem for the kink Hamiltonian in the case of a Schrodinger picture scalar field theory in 1+1 dimensions.

\subsection{The Plane Wave Decomposition}

Small perturbations about the vacuum of the free classical field theory are plane waves.  Correspondingly, the Hamiltonian of the free quantum free theory of a scalar field of mass $m$ is diagonalized by a decomposition of the Schrodinger field $\phi(x)$ and its conjugate momentum $\pi(x)$ in the plane wave basis
\beq
\phi_p=\int dx \phi(x)e^{ipx}\hsp
\pi_p=\int dx \pi(x)e^{ipx}
\eeq
which can be arranged into a basis of annihilation and creation operators
\beq
A^\dag_p=\frac{\phi_p}{2}-i\frac{\pi_p}{2\omega_p}\hsp
\frac{A_{-p}}{2\omega_p}=\frac{\phi_p}{2}+i\frac{\pi_p}{2\omega_p}\hsp
\omega_p=\sqrt{m^2+p^2}
\eeq
where the Hermitian conjugate of $A_p$ is $2\omega_p A^\dag_p$.  

One can define a plane wave normal ordering $::_a$ which places all $A$ on the right of $A^\dag$.  We remind the reader that in 1+1 dimensional scalar field theories, normal ordering is sufficient to remove all ultraviolet divergences.  In the Schrodinger picture, as fields are independent of time, such a decomposition makes no reference to the Hamiltonian and so may be performed even in an interacting theory, although it will no longer diagonalize the Hamiltonian.  

\subsection{The Kink Hamiltonian}

If the defining Hamiltonian is 
\bea
H[\pi(x),\phi(x)]&=&\int dx :\ch(\pi(x),\phi(x)):_a\\
\ch(\pi(x),\phi(x))&=&\frac{1}{2} \left(\pi^2(x)+\left(\partial_x\phi(x)\right)^2\right)+\frac{1}{g^2}V(g\phi(x))\nonumber
\eea
for a coupling constant $g$, then the kink Hamiltonian is
\beq
H\p[\pi(x),\phi(x)]=\int dx :\ch\p(\pi(x),\phi(x)):_a\hsp
\ch\p(\pi(x),\phi(x))=\ch(\pi(x),\phi(x)+f(x)). \label{hkd}
\eeq
We decompose the kink Hamiltonian into terms $H_n=\int dx \ch_n$ with $n$ factors of the fields when plane wave normal ordered and $\sum_n \ch_n=:\ch\p:_a$.  In particular
\beq
H_0=Q_0
\eeq
is the mass of the classical kink configuration $Q_0$, $H_1$ vanishes by the classical equations of motion and the free Hamiltonian density is
\beq
\ch_2(x)=\frac{1}{2}\left[
:\pi^2(x):_a+:\left(\partial_x\phi(x)\right)^2:_a+\V2 :\phi^2(x):_a
\right]
\eeq
where
\beq
\V{n}=\frac{\partial^n}{\partial(g\phi(x))^n}V(g\phi(x))|_{\phi(x)=f(x)}.
\eeq
The higher order terms are simply
\beq
\ch_{n>2}(x)=\frac{g^{n-2}}{n!}\V{n} :\phi^n(x):_a. \label{hint}
\eeq

\subsection{The Normal Mode Decomposition}

Substituting the constant frequency Ansatz
\beq
\phi(x,t)=e^{-i\omega t}\g(x)
\eeq
into the classical equations of motion derived from $H_2$ yields the wave equation
\beq
\V{2}{\g}(x)=\omega^2{\g}(x)+{\g}^{\prime\prime}(x) \label{sl}
\eeq
for the normal modes $\g(x)$.

There are three kinds of solutions.  First, there is always a zero-mode $\g_B(x)$ with $\omega_B=0$.  Second, for all real $k$ there are continuum solutions $\g_k(x)$ with $\omega_k=\sqrt{m^2+k^2}$ where $m=\sqrt{\V{2}(\pm\infty)}$.  We note that if these two limits do not agree, then the kink will accelerate \cite{tstabile,wstabile} due to a difference in the 1-loop energies of the vacua on the two sides \cite{wpol}, and so it will not correspond to any Hamiltonian eigenstate.  Finally, there may also be discrete solutions, called shape modes, $\g_S(x)$ with $0<\omega_S<m$.

For the continuum modes, we impose $\g_{-k}(x)=\g_k^*(x)$ and we impose that the discrete modes are real.  We impose that all modes are orthonormal
\beq
\int dx |{\g}_{B}(x)|^2=1,\
\int dx {\g}_{k_1} (x) {\g}^*_{k_2}(x)=2\pi \delta(k_1-k_2),\ 
\int dx {\g}_{S_1}(x){\g}_{S_2}(x)=\delta_{S_1S_2}.
\eeq
Then, as Eq.~(\ref{sl}) is a Sturm-Liouville equation, the normal modes are complete
\beq
{\g}_B(x){\g}_B(y)+\ppin{k}{\g}_k(x){\g}^*_{k}(y)=\delta(x-y) \label{comp}
\eeq
where the condensed notation $\dint$ is an integral over continuum modes plus the sum over discrete nonzero normal modes
\beq
\ppin{k}=\pin{k}+\sum_S.
\eeq

As a result of this completeness, any operator in the theory may be expanded in the normal mode basis
\beq
\phi_k=\int dx \phi(x) \g^*_k(x)\hsp
\pi_k=\int dx \pi(x) \g^*_k(x)
\eeq
where $k$ runs over all normal modes.  In the case of the zero-mode, instead of $\phi_B$ and $\pi_B$ we write $\phi_0$ and $\pi_0$.  The nonzero modes, continuous and discrete, may alternately be reexpressed in terms of Heisenberg creation and annihilation operators
\beq
B^\dag_k=\frac{\phi_k}{2}-i\frac{\pi_k}{2\omega_k}\hsp
\frac{B_{-k}}{2\omega_k}=\frac{\phi_k}{2}+i\frac{\pi_k}{2\omega_k}
\eeq
where the adjoint of $B_k$ is $2\omega_k B^\dag_k$.  Thus any operator may be expanded in the normal mode basis $\phi_0,\ \pi_0,\ B_k$ and $B^\dag_k$.  One can define {\it{normal mode normal ordering}} $::_b$ by expanding any operator in this basis and then placing all $\pi_0$ and $B_k$ on the right.  

We will assume that $f(x)$ is a BPS soliton, so that 
\beq
\int dx \left(\partial_x f(x)\right)^2=Q_0=Q_0 \int dx \g_B(x)^2.
\eeq
The zero mode $\g_B(x)$ is proportional to $\partial_x f(x)$ and so, fixing the sign of $\g_B(x)$, we conclude that
\beq
\partial_x f(x)=\sqrt{Q_0} \g_B(x). \label{fg}
\eeq

\subsection{Changing Bases}

We have seen that any Schrodinger picture operator can be decomposed in two bases.  The first is a plane wave basis defined by
\beq
[A_p,A^\dag_q]=2\pi \delta(p-q).
\eeq
The second is a normal mode basis defined by 
\beq
[B_{k_1},B^\dag_{k_2}]=2\pi \delta(k_1-k_2)\hsp
[B_S,B^\dag_S]=1\hsp [\phi_0,\pi_0]=i \label{eq:commutation}
\eeq
where for simplicity we have considered a single shape mode.  

As these bases are complete, and linear in the fields, they are related by linear Bogoliubov transformations \cite{wentzel}.  The defining Hamiltonian is plane wave normal ordered, as is the expression for the kink Hamiltonian in (\ref{hkd}).  Thus it is defined in terms of $A_p$ and $A_{-p}$.  However it will be convenient to first transform it into the $\phi_0$, $\pi_0$, $B$ and $B^\dag$ basis using the Bogoliubov transform, and then normal mode normal order it.

Normal mode normal ordering the free kink Hamiltonian, one finds \cite{cahill76,mekink}
\beq 
H_2=Q_1+\frac{\pi_0^2}{2}+\os B^\dag_SB_S+\ppin{k}\ok{} B^\dag_kB_k \label{h2p}
\eeq
where the scalar $Q_1$ is the one-loop correction to the kink mass.  Thus we find that at one-loop the center of mass motion is described by a free quantum mechanical particle with  momentum (more precisely, momentum divided by the square root of the mass $\sqrt{Q_0}$) $\pi_0$ and position (more precisely, position times $\sqrt{Q_0}$) $\phi_0$ whereas the normal modes $k$ are described by quantum harmonic oscillators with creation and annihilation operators $\Bd{}$ and $B_k$.  The ground state $\vac_0$ of this free Hamiltonian is the solution of
\beq
\pi_0\vac_0=B_k\vac_0=B_S\vac_0=0 \label{v0}
\eeq
while normal modes can be excited using $B^\dag$.  Higher order corrections to stationary states can be found \cite{me2loop} by first imposing that states are annihilated by $P\p$ and then using old fashioned perturbation theory with the interacting part of the kink Hamiltonian (\ref{hint}).

\section{Boosting a Stationary Kink} \label{boostsez}

\subsection{Copies of the Poincar\'e Algebra}

The 1+1 dimensional Poincar\'e algebra is generated by the Hamiltonian
\beq
H[\pi(x),\phi(x)]=\int dx :\ch(\pi(x),\phi(x)):_a
\eeq
the momentum operator
\beq
P[\pi(x),\phi(x)]=-\int dx :\pi(x)\partial_x\phi(x):_a
\eeq
and the boost generator 
\beq
\Lambda[\pi(x),\phi(x)]=-tP[\pi(x),\phi(x)]+\int dx x :\ch(\pi(x),\phi(x)):_a. \label{ldef}
\eeq
These generators satisfy the Poincar\'e algebra
\beq
[H,P]=0\hsp [\Lambda,H]=iP\hsp [\Lambda,P]=iH.
\eeq

Although we are in the Schrodinger picture, so that the fields do not depend on time, the boost operator has explicit time dependence when acting on a state which is not annihilated by the momentum operator $P$.  However, we will work at time $t=0$ and we will consider active transformations of the field, so that $t=0$ even after a time translation or boost.  As a result, the $-tP$ term in (\ref{ldef}) will always vanish.

Consider a state $|E,0\rangle$ such that
\beq
H|E,0\rangle=E|E,0\rangle\hsp
P|E,0\rangle=0.
\eeq
Then a boosted state
\beq
|E,\alpha\rangle=e^{-i\alpha \Lambda}|E,0\rangle
\eeq
is also an eigenvector
\beq
H |E,\alpha\rangle=E \cosh{\alpha} |E,\alpha\rangle\hsp
P |E,\alpha\rangle=E \sinh{\alpha} |E,\alpha\rangle
\eeq
identifying $\alpha$ as the rapidity of $|E,\alpha\rangle$.  In particular, for a nonrelativistic $\alpha$, the momentum of the boosted state is $E\alpha$.

In the defining Hilbert space, the time-independent states are eigenstates of $H$ and those that have fixed momentum are also eigenstates of $P$.  We have seen that these states are constructed as $\df|\psi\rangle$ where $|\psi\rangle$ is an eigenstate of $H\p$ and $P\p$.  Here $|\psi\rangle$ is found in perturbation theory.   In particular, eigenstates of $P$ with nonzero momentum are constructed by acting $\df$ on eigenstates of $P\p$ with nonzero eigenvalue.  These in turn can always be constructed from eigenstates of $P\p$ with zero eigenvalue by acting with a boost $\Lambda\p$ defined by
\beq
\Lambda\p=\df^\dag \Lambda\df
\eeq
as the kink operators satisfy another copy of the Poincar\'e algebra
\beq
[H\p,P\p]=0\hsp [\Lambda\p,H\p]=iP\p\hsp [\Lambda\p,P\p]=iH\p.
\eeq

If
\beq
H\p|E,0\rangle=E|E,0\rangle\hsp
P\p|E,0\rangle=0
\eeq
then
\beq
H\p e^{-i\alpha \Lambda\p}|E,0\rangle=E \cosh{\alpha} e^{-i\alpha \Lambda\p}|E,0\rangle\hsp
P\p e^{-i\alpha \Lambda\p}|E,0\rangle=E \sinh{\alpha} e^{-i\alpha \Lambda\p}|E,0\rangle
\eeq
and so $e^{-i\alpha\Lambda\p}$ boosts a state annihilated by $P\p$ to one with eigenvalue $E\alpha$ if $\alpha<<1$.   

Therefore our strategy will be as follows.  We begin with an eigenstate $|\Psi\rangle$ of $H\p$ which is annihilated by $P\p$, constructed as described in Sec.~\ref{revsez}.  This corresponds, in the defining Hilbert space, to a state $\df|\Psi\rangle$ which is annihilated by $P$, a stationary kink.  Then
\beq
e^{-i\alpha\Lambda}\df|\Psi\rangle =\df e^{-i\alpha\Lambda\p}|\Psi\rangle \label{boostato}
\eeq
is our desired eigenstate of $H$ with rapidity $\alpha$.  Thus we will have constructed a kink state with nonzero momentum.   The right hand side of Eq.~(\ref{boostato}) is our first construction of a boosted kink state.  We will spend the rest of this section trying to understand it.

\subsection{The Kink Boost Operator}

In this subsection we will calculate $\Lambda\p$, and expand it order by order in our semiclassical expansion.

For any functional $:F[\pi(x),\phi(x)]:$ with any normal ordering prescription \cite{mekink}
\beq
:F[\pi(x),\phi(x)]:\df=\df :F[\pi(x),\phi(x)+f(x)]:.
\eeq
Therefore the kink momentum is
\bea
P\p[\pi(x),\phi(x)]&=&P[\pi(x),\phi(x)+f(x)]\\
&=&-\int dx :\pi(x)\partial_x\phi(x):_a-\int dx \pi(x) \partial_x f(x)=P[\pi(x),\phi(x)]-\sqrt{Q_0}\pi_0\nonumber
\eea
where in the last step we have used Eq.~(\ref{fg}).  Similarly the kink boost operator is
\bea
\Lambda\p[\pi(x),\phi(x)]&=&\df^\dag \Lambda[\pi(x),\phi(x)] \df=\Lambda[\pi(x),\phi(x)+f(x)]\\
&=&\int dx x :\ch(\pi(x),\phi(x)+f(x)):_a=\int dx x :\ch\p(\pi(x),\phi(x)):_a\nonumber\\
&=&\int dx x \left[\frac{1}{2} \left(:\pi^2(x):_a+:\left(\partial_x\left(\phi(x)+f(x)\right)\right)^2:_a\right)\right.\nonumber\\
&&\left.+\frac{1}{g^2}:V(g\phi(x)+gf(x)):_a\right].\nonumber
\eea

Let us expand this order by order in the fields $\phi(x)$ and $\pi(x)$
\beq
\Lambda\p=\sum_n \Lambda_n\p.
\eeq
At zeroeth order, for symmetric solutions $f(x)=f(-x)$, one obtains
\beq
\Lambda\p_0=\int dx x \left[\frac{1}{2} \left(\partial_xf(x)\right)^2+\frac{1}{g^2}V(gf(x))\right]=0
\eeq
which vanishes as $x$ is odd and the term in parenthesis is even.  Here we ignore the linear divergence at large $|x|$, which can be eliminated by shifting the potential by a constant so that $V$ vanishes at the vacua $gf(\pm\infty)$.  This is anyway achieved by the infrared counterterms included in this approach \cite{mephi4}.

At first order
\bea
\Lambda\p_1&=&\int dx x \left[(\partial_x\phi(x))(\partial_xf(x))+\frac{\phi(x)}{g}V\p(gf(x))\right]\nonumber\\
&=&\int dx \phi(x)\left[- \partial_x\left(x\partial_xf(x)\right)+\frac{1}{g}\V1\right]\nonumber\\
&=&-\int dx \phi(x)\partial_x f =-\sqrt{Q_0}\phi_0
\eea
where, going from the second to the third line, we used the classical equations of motion satisfied by $f(x)$ and on the last line we used (\ref{fg}).  The classical kink mass $Q_0$ is of order $m/g^2$ and so the coefficient $\sqrt{Q_0}$ is of order $\sqrt{m}/g$.  

The quadratic terms are
\bea
\Lambda\p_2&=&\int dx \frac{x}{2} :\left[\pi^2(x)+\left(\partial_x \phi(x)\right)^2+\phi^2(x)\V2 \right]:_a\\
&=&\int dx \frac{x}{2} :\left[\pi^2(x) +\phi(x) \left(-\partial^2_x \phi(x)+\V2\phi(x)\right) \right]:_a-\frac{1}{2}\int dx :\phi(x)\partial_x \phi(x):_a\nonumber
\eea
where the last term is a total derivative which vanishes if $\phi^2(\infty)=\phi^2(-\infty)$, which we will impose, thus dropping the boundary terms from our boost operator.  Using the decompositions
\beq
\phi(x)=\phi_0 \g_B(x) + \ppin{k} \phi_k \g_k(x)\hsp
\pi(x)=\pi_0 \g_B(x) + \ppin{k} \pi_k \g_k(x) \label{pdec}
\eeq
and (\ref{sl}) one can simplify the term in parenthesis
\beq
\Lambda\p_2=\int dx \frac{x}{2} :\left[\pi^2(x) +\phi(x) \ppin{k}\phi_k\omega_k^2 \g_k(x)\right]:_a.
\eeq
In terms of $\Delta$ symbols, defined in (\ref{deldef}), this is
\bea
\Lambda\p_2&=&\ppink{2}\frac{\Delta^{100}_{k_1k_2}}{2}:\left(\pi_{k_1}\pi_{k_2}+\omega_{k_1}^2 \phi_{k_1}\phi_{k_2}\right):_a+\ppin{k}\Delta^{100}_{B k}:\left(\pi_{0}\pi_{k}+\frac{\omega_{k}^2}{2} \phi_{0}\phi_{k}\right):_a\\
&=&\ppink{2}\frac{\Delta^{001}_{k_1k_2}}{\omega_{k_2}^2-\omega_{k_1}^2}:\left(\pi_{k_1}\pi_{k_2}+\omega_{k_1}^2\phi_{k_1}\phi_{k_2}\right):_a+\ppin{k}\Delta^{001}_{B k}:\left(\frac{2}{\omega^2_k}\pi_{0}\pi_{k}+ \phi_{0}\phi_{k}\right):_a\nonumber
\eea
where we used the fact that for a symmetric kink $\Delta^{100}_{BB}$ vanishes and (\ref{did}).  To simplify things later, we will change plane wave normal ordering to normal mode normal ordering.  This shifts $\Lambda\p_2$ by a real number, and so it shifts the translation operator $e^{-i\alpha\Lambda\p}$ by a phase.  As the total phase of the state is not measurable, we simply drop this constant, leaving
\beq
\Lambda\p_2=\ppink{2}\frac{\Delta^{001}_{k_1k_2}}{\omega_{k_2}^2-\omega_{k_1}^2}:\left(\pi_{k_1}\pi_{k_2}+\omega_{k_1}^2\phi_{k_1}\phi_{k_2}\right):_b+\ppin{k}\Delta^{001}_{B k}\left(\frac{2}{\omega^2_k}\pi_{0}\pi_{k}+ \phi_{0}\phi_{k}\right). \label{l2}
\eeq
Note that no normal ordering is needed on the last term as $\phi_0$ and $\pi_0$ both commute with $B^\dag$ and $B$, so the normal mode normal ordering does nothing.

The higher order terms are
\beq
\Lambda\p_{n>2}=\frac{g^{n-2}}{n!}\int dx x :\phi^n(x):_a\V{n}.
\eeq
Again these may be expanded into $\phi_0$, $\pi_0$, $\phi_k$ and $\pi_k$ using (\ref{pdec}).

\subsection{The Moduli Space Coordinate}

Recall that a rapidity $\alpha$ boost is achieved with the operator $e^{-i\alpha\Lambda\p}$. For concreteness, let us consider the kink ground state $\vac$ written, in the kink Hilbert space, as an eigenvector of $H\p$ with $H\p\vac=Q\vac$.  Then the corresponding boosted state, still working in the kink Hilbert space\footnote{Recall that the action of $\df$ takes this state to the defining Hilbert space.}, is
\beq
|\alpha\rangle=e^{-i\alpha\Lambda\p}\vac. \label{boo}
\eeq

In the rest of this section we will evaluate (\ref{boo}) one order at a time. In Subsec.~\ref{1sez} we will truncate the kink ground state $\vac$ to the one-loop kink ground state $\vac_0$ which satisfies (\ref{v0}).

Our first task is to write this state in a convenient basis.  Recall from (\ref{eq:commutation}) that our operator algebra is the product of a commuting quantum mechanical canonical algebra generated by $\pi_0$ and $\phi_0$ with an infinite set of Heisenberg algebras $B_k$ and $B^\dag_k$, with $k$ running over all real numbers and possibly some discrete values corresponding to shape modes.  Therefore the Hilbert space factorizes into the product of the Harmonic oscillator Fock spaces for each $k$ with the space of quantum mechanical wave functions which form a representation of $\pi_0$ and $\phi_0$.  These wave functions are defined by
\beq
|\psi\rangle=\int dy \psi(y)|y\rangle\hsp \phi_0|\psi\rangle=\int dy y\psi(y)|y\rangle\hsp
 \pi_0|\psi\rangle=-i\int dy \frac{\partial\psi(y)}{\partial y}|y\rangle.
\eeq
So to describe a state, for each element of the harmonic oscillator Fock space, one needs a complex wave function $\psi(y)$.

The one-loop ground state, which solves (\ref{v0}), is easy to write in this basis.  Let $|y\rangle_0$ be the Fock space element annihilated by all operators $B_k$
\beq
B_k|y\rangle_0=0\hsp \phi_0|y\rangle_0=y|y\rangle_0
\eeq
and choose the function $\psi(y)$ to be a constant
\beq
\vac_0=\int dy |y\rangle_0.
\eeq
The choice of constant is just a normalization convention, although these states are nonnormalizable.

We will systematically investigate all of the perturbative expansions involved in our construction.  Let us begin with the unboosted one-loop ground state $\vac_0$ itself.  This is found using perturbation theory, which produces corrections of the form $mg\phi_0^2$ in the semiclassical expansion.  Acting on our basis, the semiclassical expansion is therefore a series in $mgy^2$.  Therefore the one-loop ground state $\vac_0$ itself is only a good approximation to the ground state at
\beq
y<<\frac{1}{\sqrt{mg}}. \label{ymax}
\eeq
Of course since $\psi(y)$ is a constant, the wave function is supported at all values of $y$, including those not satisfying (\ref{ymax}).  Thus one should not trust the perturbative expansion on that part of the wave function.  

The situation is similar to solving for a bound wave function in quantum mechanics as a power series in the space coordinate $x$.  The wave function in that case is reliable only for small $x$.  

What is $y$ physically?  Let us compute the scalar field profile corresponding to the state $|y\rangle_0$, shifted back to the defining Hilbert space using $\df$
\bea
\frac{{}_0\langle y|\df^\dag \phi(x)\df|y\rangle_0}{{}_0\langle y|\df^\dag \df|y\rangle_0}&=&\frac{{}_0\langle y|\phi(x)+f(x)|y\rangle_0}{{}_0\langle y|y\rangle_0}=f(x)+\frac{{}_0\langle y|\phi_0 \g_B(x)|y\rangle_0}{{}_0\langle y|y\rangle_0}\\
&=&f(x)+y \g_B(x)=f(x)+\frac{y}{\sqrt{Q_0}}\partial_x f(x)=f\left(x+\frac{y}{\sqrt{Q_0}}\right)+O(y^2).\nonumber
\eea
Recall that there is a moduli space of kink solutions $f(x-x_0)$ related by a spatial translation $x_0$.  The parameter $y$ is a coordinate on this moduli space, and
\beq
x_0=-y/\sqrt{Q_0} \label{x0}
\eeq
is the translation.  It is thus reasonable that a zero-momentum kink has a wave function $\psi(y)$ which is independent of $y$, as it is translation-invariant.

Now we may interpret the expansion in $mgy^2$.  As $Q_0\sim m/g^2$ and $y$ is proportional to the kink position $x_0$ times $\sqrt{Q_0}$, this is an expansion in $mgQ_0 x_0^2\sim m^2x_0^2/g$.  So this is an expansion in the distance $x_0$ to the center of mass of the kink, with convergence in the sense of an asymptotic series when the kink position $x_0$ varies by less than $\sqrt{g}/m$.  Here $1/m$ is the size of the classical kink solution itself.  This condition is physically reasonable, the semiclassical approximation implies that the kink is, by at least a factor of $\sqrt{g}$, more localized than the size of the solution itself, so that the solution is not too smeared by quantum effects.

\subsection{Boosting the One-Loop Kink} \label{1sez}

In this subsection we will boost the one-loop kink ground state, evaluating
\beq
e^{-i\alpha\Lambda\p}\vac_0
\eeq
in perturbation theory.  We start with the leading order contribution
\beq
|\alpha\rangle_0=
e^{-i\alpha\Lambda_1\p}\vac_0=e^{i\sqrt{Q_0}\alpha\phi_0}\vac_0=\int dy e^{i\sqrt{Q_0}\alpha y}|y\rangle_0. \label{al0}
\eeq
Alternately this state may be defined by 
\beq
B_k|\alpha\rangle_0=0\hsp \pi_0|\alpha\rangle_0=\sqrt{Q_0}\alpha|\alpha\rangle_0.
\eeq

Using (\ref{x0}), the phase in the wave function (\ref{al0}) may be written
\beq
e^{i\sqrt{Q_0}\alpha y}=e^{-i Q_0\alpha x_0}.
\eeq
This phase is of the usual plane wave form $e^{-ipx_0}$ where the momentum $p$ is identified with $Q_0\alpha$.  At low rapidity, $\alpha$ is simply the velocity $v$ and at leading order in the semiclassical expansion, $Q_0$ is the mass $M$ and so this is just the Newtonian formula $p=Mv$ for the momentum.

Now let us try to include the next order correction to the boost operator $\Lambda$.  Consider
\beq
e^{-i\alpha\left(\Lambda_1\p+\Lambda_2\p\right)}\vac_0=e^{i\alpha\left(\sqrt{Q_0}\phi_0-\Lambda_2\p\right)}\vac_0 \label{l12}
\eeq
where $\Lambda\p_2$ is given in (\ref{l2}).  The exponential consists of quadratic and linear terms in the fields, and so it acts as a Bogoliubov transformation.  Physically, it ensures that the boosted state, at this order, is annihilated not by the normal mode annihilation operators $B_k$, but rather by the annihilation operators corresponding to boosted normal modes.  In practice, finding these boosted normal modes suffices for calculating the action of various operators on the boosted state.

On the other hand, expressing this state in terms of $\vac_0$ is quite complicated.  The problem is that $\Lambda\p_1$ and $\Lambda\p_2$ do not commute, and their commutator does not commute with $\Lambda\p_2$.  This series of commutators does not truncate.   

The first term in the series consists of terms in which $\Lambda\p_2$ does not appear.  This is the state $|\alpha\rangle_0$ given in (\ref{al0}).  We will now calculate the subleading correction, in which $\Lambda\p_2$ appears once in the exponential.  First, note that the only term in $\Lambda\p_2$  which does not commute with $\Lambda\p_1$ is $\pi_0 \dint\frac{dk}{2\pi}\Delta^{001}_{B k}\frac{2}{\omega^2_k}\pi_{k}$.  So first let us include only that term, using the Baker Campbell Hausdorff formula
\bea
&&\hspace{-1cm}\exp{i\alpha\left(\sqrt{Q_0}\phi_0-\pi_0 \ppin{k}\Delta^{001}_{B k}\frac{2}{\omega^2_k}\pi_{k}\right)}\vac_0\label{bch}\\
&&=\exp{i\alpha\sqrt{Q_0}\phi_0}\exp{-i\alpha\pi_0 \ppin{k}\Delta^{001}_{B k}\frac{2}{\omega^2_k}\pi_{k}}\nonumber\\
&&\times\exp{-\frac{1}{2}\left[i\alpha\sqrt{Q_0}\phi_0,-i\alpha\pi_0 \ppin{k}\Delta^{001}_{B k}\frac{2}{\omega^2_k}\pi_{k}\right]}\vac_0\nonumber\\
&&=\exp{i\alpha\sqrt{Q_0}\phi_0}\exp{-\left[i\alpha\sqrt{Q_0}\phi_0,-i\alpha\pi_0 \ppin{k}\Delta^{001}_{B k}\frac{1}{\omega^2_k}\pi_{k}\right]}\vac_0\nonumber\\
&&=\exp{i\alpha\sqrt{Q_0}\phi_0}\exp{-i\alpha^2\sqrt{Q_0}\ppin{k}\frac{\Delta^{001}_{B k}}{\omega^2_k}\pi_{k}}\vac_0\nonumber\\
&&=\exp{\alpha^2\sqrt{Q_0} \ppin{k}\frac{\Delta^{001}_{B k}}{\omega_k}B_k^\dag}|\alpha\rangle_0\nonumber\\
&&=\left(1+\alpha^2\sqrt{Q_0} \ppin{k}\frac{\Delta^{001}_{B k}}{\omega_k}B_k^\dag+O\left(\frac{\alpha^4}{g^2}\right)\right)|\alpha\rangle_0.\nonumber
\eea
This is an expansion in $\alpha^2/g$, and so it is expected to converge when $\alpha^2<<g$.  This means for example that the kink kinetic energy, which nonrelativistically is of order $Q\alpha^2\sim m\alpha^2/g^2$, should be less than $Qg\sim m/g$.  The kink kinetic energy may be much larger than the meson mass $m$, but still this expansion is only valid in the deep nonrelativistic regime.  Similarly the kink momentum $Q\alpha\sim m\alpha/g^2$ should be less than $m/g^{3/2}$.  

Including the other terms in (\ref{l2}), again at linear order in $\Lambda\p_2$, the boosted state (\ref{l12}) becomes
\bea
&&\left[1+\alpha^2\sqrt{Q_0} \ppin{k}\frac{\Delta^{001}_{B k}}{\omega_k}B_k^\dag\right.\label{d12f}\nonumber\\
&&\left. -i\alpha\left(
\ppink{2}\frac{\Delta^{001}_{k_1k_2}}{\omega_{k_2}^2-\omega_{k_1}^2}:\left(\pi_{k_1}\pi_{k_2}+\frac{\omega_{k_1}^2+\omega_{k_2}^2}{2}\phi_{k_1}\phi_{k_2}\right):_b+\ppin{k}\Delta^{001}_{B k} \phi_{0}\phi_{k}
\right)\right]|\alpha\rangle_0\nonumber\\
&=&\left[1+\alpha^2\sqrt{Q_0} \ppin{k}\frac{\Delta^{001}_{B k}}{\omega_k}B_k^\dag\right.\nonumber\\
&&+\left. i\alpha\left(
\ppink{2}\frac{\Delta^{001}_{k_1k_2}}{2}\frac{\omega_{k_1}-\omega_{k_2}}{\omega_{k_1}+\omega_{k_2}}B_{k_1}^\dag B^\dag_{k_2}-\phi_{0}\ppin{k}\Delta^{001}_{B k} B_k^\dag
\right)\right]|\alpha\rangle_0\nonumber.
\eea
The additional terms are the first terms in a series in $\alpha$, which is convergent whenever $\alpha<<1$.  Thus this requires the kink to be nonrelativistic.  As $g<<1$, this bound is weaker than the bound required for the convergence of the series in Eq.~(\ref{bch}), and so it does not represent a new constraint on the validity of our approximation.

The interaction terms $\Lambda\p_{n>2}$ all commute with $\Lambda\p_1$ but not with $\Lambda\p_2$, and so they can also be pulled out of the expression (\ref{al0}) for $\alpha_0$.  The plane wave normal ordering of these terms is easily converted to normal mode normal ordering using the Wick's theorem of Ref.~\cite{mewick}.  They therefore simply add terms to the left hand side of (\ref{d12f}) that are cubic and higher in $\phi_0$ and $B^\dag$.  For example, the cubic term yields a factor of
\beq
-i\frac{\alpha g}{6}\int dx x \V{n}\left(:\phi^3(x):_b+6\I(x)\phi(x)\right)
\eeq
where
$\I(x)$ is 
\beq
\I(x)=\pin{k}\frac{\left|{\g}_{k}(x)\right|^2-1}{2\omega_k}+\sum_S \frac{\left|{\g}_{S}(x)\right|^2}{2\omega_k}.\label{di}
\eeq
Acting on $|\alpha\rangle_0$ one may drop the annihilation operators, leaving the contribution
\bea
|\alpha\rangle&\supset&-i\alpha g\ppin{k}\left(\int dx x \V{n}\I(x)\g_{k}(x)\right)B^\dag_{k}|\alpha\rangle_0\\
&&-i\frac{\alpha g}{6}\ppink{3}\left(\int dx x \V{n}\g_{k_1}(x)\g_{k_2}(x)\g_{k_3}(x)\right)B^\dag_{k_1}B^\dag_{k_2}B^\dag_{k_3}|\alpha\rangle_0\nonumber
\eea
plus terms where each subset of the $k$ is replaced by zero modes, so that the corresponding $\g_k$ all become $\g_B$ and $B^\dag_k$ become $\phi_0$.


\subsection{Boosting the Next Order Kink} \label{15sez}

At next order in $g$, the vacuum $\vac_1$ consists of four terms, proportional to $\phi_0^2B^\dag_{k_1}\vac_0$, $\phi_0B^\dag_{k_1}B^\dag_{k_2}\vac_0$, $B^\dag_{k_1}\vac_0$ and $B^\dag_{k_1}B^\dag_{k_2}B^\dag_{k_3}\vac_0$.  The first two are universal, in the sense that they are entirely fixed by the translation invariance of $\df\vac$.  The other two depend on the precise form of the potential $V$.  Let us consider here only the universal terms
\beq
\vac_1=\frac{Q_0^{-1/2}}{2}\ppin{k_1}\omega_{k_1}\Delta^{001}_{k_1 B}\phi_0^2B^\dag_{k_1}\vac_0+Q_0^{-1/2}\ppink{2}\omega_{k_1}\Delta^{001}_{k_1k_2}\phi_0B^\dag_{k_1}B^\dag_{k_2}\vac_0.
\eeq

The leading order boost is
\bea
e^{-i\alpha\Lambda_1\p}\vac_1&=&\frac{Q_0^{-1/2}}{2}\ppin{k_1}\omega_{k_1}\Delta^{001}_{k_1 B}\phi_0^2B^\dag_{k_1}\int dy y^2 e^{i\sqrt{Q_0}\alpha y}|y\rangle_0\\
&&+Q_0^{-1/2}\ppink{2}\omega_{k_1}\Delta^{001}_{k_1k_2}B^\dag_{k_1}B^\dag_{k_2}\int dy y e^{i\sqrt{Q_0}\alpha y}|y\rangle_0.\nonumber
\eea
Including the 1-loop ground state this is
\bea
e^{-i\alpha\Lambda_1\p}\left(\vac_0+\vac_1\right)&=&\int dy \left(1+y^2\frac{Q_0^{-1/2}}{2}\ppin{k_1}\omega_{k_1}\Delta^{001}_{k_1 B}B^\dag_{k_1}\right. \label{l1s}\\
&&\left.+yQ_0^{-1/2}\ppink{2}\omega_{k_1}\Delta^{001}_{k_1k_2}B^\dag_{k_1}B^\dag_{k_2}
\right)e^{i\sqrt{Q_0}\alpha y}|y\rangle_0.\nonumber
\eea

We cannot yet calculate form factors, because our states are nonnormalizable, being momentum eigenstates.  In Sec.~\ref{pacsez} we will introduce wave packets states, whose form factors will be calculated in a companion paper.  However, ignoring this problem for a moment, one leading contribution to the naive form factor $ \langle 0|\df^\dag \phi(x)\df|\alpha\rangle$, after the classical contribution equal to $f(x)$ times the normalization of the state, arises from ${}_0\langle 0|\df^\dag \phi(x)\df$ acting on the last term in (\ref{l1s})
\bea
&& e^{-i\alpha(\Lambda_1\p+\Lambda_2\p)}\vac_1\supset-i\alpha\Lambda\p_2 e^{-i\alpha\Lambda_1\p}\vac_1\\
&&\supset-i\alpha\Lambda\p_2\ppin{k\p}\Delta^{001}_{B k\p}\frac{2}{\omega^2_{k\p}}\pi_{0}\pi_{k\p}\int dy \phi_0 Q_0^{-1/2}\ppink{2}\frac{\omega_{k_1}-\omega_{k_2}}{2}\Delta^{001}_{k_1k_2}B^\dag_{k_1}B^\dag_{k_2}
e^{i\sqrt{Q_0}\alpha y}|y\rangle_0\nonumber\\
&&\supset \frac{\alpha}{\sqrt{Q_0}}\ppink{2}\left(\omega_{k_2}-\omega_{k_1}\right)\frac{\Delta^{001}_{B-k_2}}{\omega_{k_2}^2}\Delta^{001}_{k_1k_2}B^\dag_{k_1}
|\alpha\rangle_0 \nonumber\\
&&=\frac{\alpha}{2\sqrt{Q_0}}\ppink{2}\left(\omega_{k_2}-\omega_{k_1}\right)\Delta^{100}_{B-k_2}\Delta^{001}_{k_1k_2}B^\dag_{k_1}
|\alpha\rangle_0.\nonumber
\eea
The other contributions, arising from $ {}_1\langle 0|\df^\dag \phi(x)\df|\alpha\rangle_0$ and from the $\Lambda\p_3$ term in $ {}_0\langle 0|\df^\dag \phi(x)\df|\alpha\rangle_0$, can be computed similarly.  

\section{A Normalizable Wave Packet} \label{pacsez}

\subsection{Two Kinds of Wave Packets}

Sec.~\ref{boostsez} describes momentum eigenstates.  These are solitons whose wave packets are very delocalized with respect to their size and so are effectively plane waves.  In this section we turn our attention to soliton wave packets that are narrower than the soliton size, so that the quantum profile is well approximated by the classical profile.  In particular, since the solitons are spatially limited, the soliton states themselves will be normalizable.  This will allow us to define and to calculate, for the first time using linearized soliton perturbation theory, matrix elements of soliton states.

\subsection{The Simplest Wave Packet}

Unlike the momentum eigenstates of Sec.~\ref{boostsez}, localized wave packets are not unique, not even after specifying a finite number of quantum numbers.   Also, unlike those, they will be neither Hamiltonian nor momentum eigenstates.  Thus this construction is somewhat arbitrary.  One may try to make the states as close to Hamiltonian eigenstates as possible, but whether that corresponds to the physical state describing some specific soliton depends on its history.  

We will therefore choose two somewhat arbitrary criteria for our states.  First, they should be as simple as possible.  Second, they should be sufficiently localized that our perturbation theory converges in the sense of an asymptotic series.  In other words, the eigenvalue $y$ of $\phi_0$ should be supported in a region satisfying (\ref{ymax}), which implies in particular that the wave packet width should be smaller than the inverse meson width, which itself is roughly the size of the classical soliton solution.

This motivates the following choice
\beq
\as=\frac{1}{(2\pi)^{1/4}\sqrt{\sigma}}e^{-\frac{\phi_0^2}{4\sigma^2}}|\alpha\rangle_0. \label{sig}
\eeq
Of course, one may replace $|\alpha\rangle_0$ by a better approximation to $|\alpha\rangle$ to obtain something closer to a momentum or Hamiltonian eigenstate.  For example one could include more quantum corrections.  But we will not do this here.  Intuitively, the fact that we use $\vac_0$ and drop $\vac_1$ in our construction, implies that the kink center of mass has momentum but it is not correlated to that of its normal mode cloud.  

Here we are, as always, working in the kink Hilbert space obtained by acting on the defining Hilbert space with $\df^\dag$.  Thus, in the defining Hilbert space, our wave packet is
\beq
\df\as.
\eeq

We will fix our normalization using the convention
\beq
{}_0\langle y_1|y_2\rangle_0=\delta(y_1-y_2).
\eeq
Inserting (\ref{al0}) into (\ref{sig}) one finds
\beq
\as=\frac{1}{(2\pi)^{1/4}\sqrt{\sigma}}\int dy \exp{-\frac{y^2}{4\sigma^2}+i\sqrt{Q_0}\alpha y}|y\rangle_0. 
\eeq
In particular, the wave packet is normalized to unity
\beq
\asb\df^\dag\df\as=\asb 1\as=\frac{1}{\sigma\sqrt{2\pi}}\int dy \exp{-\frac{y^2}{2\sigma^2}}=1.
\eeq

\subsection{Matrix Elements}

The main result of the present note is that matrix elements of kink wave packets are easy to compute using our formalism.   Such matrix elements have applications to many physical processes of interest, such as calculating the probability to excite a shape mode during kink-meson scattering, the calculation of form factors, kink-impurity scattering, etc.  In the present note we will calculate only those matrix elements which are necessary to understand the wave packet itself and to show which range of $\sigma$ and $\alpha$ is simultaneously compatible with the perturbative expansion (\ref{ymax}) and also allows the kink rapidity to be localized near $\alpha$.  We will not consider applications to specific physical processes.

\subsubsection{The Kink Position}

First, let us try to understand the meaning of $\sigma$ by computing matrix elements of $\phi_0$.  Note that
\beq
\asb\phi_0\as=\frac{1}{\sigma\sqrt{2\pi}}\int dy \exp{-\frac{y^2}{2\sigma^2}}y=0
\eeq
and so this wave packet is centered at $y=0$.  Recalling (\ref{x0}) this implies that the kink is centered at the base point $x_0=0$.  To evaluate its smearing, one calculates
\beq
\asb\phi_0^2\as=\frac{1}{\sigma\sqrt{2\pi}}\int dy \exp{-\frac{y^2}{2\sigma^2}}y^2=\sigma^2.
\eeq
Thus one sees that $y$ has a variance of $\sigma^2$ and a standard deviation of $\sigma$.  Using (\ref{x0}) one sees that $x_0$ has a standard deviation of
\beq
\sigma_{x_0}=\frac{\sigma}{\sqrt{Q_0}}.
\eeq
Thus $\sigma$ characterizes the coherent spatial smearing of the kink wave packet.  Recalling that the classical solution has a width of $1/m$, the semiclassical condition $\sigma_{x_0}<<1/m$ that the quantum smearing is smaller than the classical length scale is equivalent to
\beq
\sigma<<\frac{\sqrt{Q_0}}{m}\sim\frac{1}{\sqrt{m}{g}}. \label{siglim}
\eeq

Note that this is weaker than the condition (\ref{ymax}) that our perturbation series converges.  The perturbation series is an expansion in, among other things, $mg\phi_0^2$ and so it converges when
\beq
\sigma<<\frac{1}{\sqrt{mg}}. \label{pertlim}
\eeq

\subsubsection{The Kink Momentum}

Let us begin with
\beq
\asb\pi_0\as=\frac{-i}{\sigma\sqrt{2\pi}}\int dy \exp{-\frac{y^2}{2\sigma^2}}\left(-\frac{y}{2\sigma^2}+i\sqrt{Q_0}\alpha\right)=\sqrt{Q_0}\alpha.
\eeq
Thus the expected momentum contained in the kink center of mass is
\beq
\asb-\sqrt{Q_0}\pi_0\as=Q_0\alpha.
\eeq
This is just the leading order product of the mass times the velocity, as expected for the nonrelativistic momentum.

The momentum contained in the normal modes is described by the momentum operator~\cite{me2loop}
\beq
P=-\int dx :\pi(x)\partial_x\phi(x):_a=\ppink{2}:\phi_{k_1}\pi_{k_2}:_b\Delta^{001}_{k_1k_2}+\pi_0\ppin{k}\phi_k \Delta^{001}_{kB}-\phi_0\ppin{k}\pi_k \Delta^{001}_{kB}. \label{pb}
\eeq
As a result of the normal mode normal ordering in the last expression,
\beq
{}_0\langle y_1|P|y_2\rangle_0=0
\eeq
and so
\beq
\asb P\as=0.
\eeq
Physically, this means that the normal modes do not carry any momentum in the state $|\sigma\rangle$.  Similarly, as a result of the $B$ and $B^\dag$ in each term in Eq.~(\ref{pb}),
\beq
\asb P\pi_0\as=\asb \pi_0 P\as=0.
\eeq

The total momentum carried by the wave packet is
\beq
\asb\df^\dag P\df\as=\asb (P-\sqrt{Q_0}\pi_0)\as=Q_0\alpha \label{pvev}
\eeq
which again agrees with the nonrelativistic expression.  So the wave packet state $\df\as$ indeed has its momentum peaked about the desired value.  

\subsubsection{The Kink Momentum Spread}

The variance of the momentum is
\beq
\asb\df^\dag P^2\df\as-\left(\asb\df^\dag P\df\as\right)^2=\asb \left(P-\sqrt{Q_0}\pi_0\right)^2\as - Q_0^2 \alpha^2. \label{varinit}
\eeq

To claim that $\df\as$ is a good approximation to a momentum eigenstate, at least for some range of $\alpha$ and $\sigma$, the standard deviation of the momentum should be less than its expectation value.  Let us next check this.  First, note that
\beq
\asb Q_0 \pi^2_0\as=-Q_0{\sigma\sqrt{2\pi}}\int dy \exp{-\frac{y^2}{2\sigma^2}}\left[\left(-\frac{y}{2\sigma^2}+i\sqrt{Q_0}\alpha\right)^2-\frac{1}{2\sigma^2}\right]=\frac{Q_0}{4\sigma^2}+Q^2_0\alpha^2. 
\eeq
The last term cancels the last term in (\ref{varinit}), leaving a contribution to the variance of $Q_0/(4\sigma^2)$.

Let us check this result against the uncertainty principle.   The kink center of mass has been localized to a spatial distance of $\sigma/\sqrt{Q_0}$, leading to a momentum standard deviation of order $O(\sqrt{Q_0}/\sigma)$.  This indeed is the square root of the above contribution to the variance.

When the semiclassical approximation (\ref{siglim}) holds, the corresponding momentum uncertainty is
\beq
\sqrt{\asb Q_0 \pi^2_0\as}=\sqrt{\frac{Q_0}{4\sigma^2}}>>\frac{m}{2}. \label{lower}
\eeq
In other words, the kink center of mass momentum spread is at least the meson mass.  This means that our wave packet will only be useful for processes involving relativistic mesons.

There is one more contribution to the momentum spread, arising from the kink's normal mode cloud
\bea
\asb P^2\as&=&\ppink{4}\Delta^{001}_{k_1 k_2}\Delta^{001}_{k_3 k_4}\asb :\phi_{k_1}\pi_{k_2}:_b :\phi_{k_3}\pi_{k_4}:_b \as\nonumber\\
&&+ \ppink{2}\Delta^{001}_{k_1 B}\Delta^{001}_{k_2 B}\asb \phi_0^2 \pi_{k_1}\pi_{k_2}\as\nonumber\\
&&- \ppink{2}\Delta^{001}_{k_1 B}\Delta^{001}_{k_2 B}\left(\asb \phi_0 \pi_0 \phi_{k_1}\pi_{k_2}\as+\asb \pi_0 \phi_0 \pi_{k_1}\phi_{k_2}\as\right)\nonumber\\
&&+ \ppink{2}\Delta^{001}_{k_1 B}\Delta^{001}_{k_2 B}\asb \pi_0^2 \phi_{k_1}\phi_{k_2}\as . \label{p2}
\eea
Note that
\bea
i+\asb \pi_0 \phi_0 \as&=&\asb \phi_0 \pi_0 \as\\
&=&\frac{-i}{\sigma\sqrt{2\pi}}\int dy \exp{-\frac{y^2}{2\sigma^2}}y\left(-\frac{y}{2\sigma^2}+i\sqrt{Q_0}\alpha\right)=\frac{i}{2}.\nonumber
\eea
Therefore the matrix elements are
\bea
\asb :\phi_{k_1}\pi_{k_2}:_b :\phi_{k_3}\pi_{k_4}:_b \as&=&
\asb \frac{B_{-k_1}}{2\ok{1}} \frac{B_{-k_2}}{2}\Bd3\ok4\Bd4 \as\nonumber\\
&&\hspace{-4cm}=\frac{\ok4}{4\ok1}(2\pi)^2\left(\delta(k_1+k_3)\delta(k_2+k_4)+\delta(k_1+k_4)\delta(k_2+k_3)\right)
\eea
and
\bea
\asb \phi_0^2 \pi_{k_1}\pi_{k_2}\as&=&\frac{\ok2}{2}\asb \phi_0^2 B_{-k_1}\Bd2\as=\ok2\sigma^2\pi\delta(k_1+k_2)\\
\asb \pi_0^2 \phi_{k_1}\phi_{k_2}\as&=&\frac{1}{2\ok1}\asb \pi_0^2 B_{-k_1}\Bd2\as=\left(\frac{1}{4\sigma^2}+Q_0\alpha^2\right)\frac{\pi\delta(k_1+k_2)}{\ok1}\nonumber
\eea
and finally
\bea
\asb \phi_0 \pi_0 \phi_{k_1}\pi_{k_2}\as&=&\frac{i\ok2}{2}\frac{i}{2\ok1}\asb B_{k_1}\Bd2\as=-\frac{\pi\delta(k_1+k_2)}{2}\label{croce}\\
\asb \pi_0 \phi_0 \pi_{k_1}\phi_{k_2}\as&=&\left(\frac{-i}{2}
\right)\left(\frac{-i}{2}\right)\asb B_{k_1}\Bd2\as=-\frac{\pi\delta(k_1+k_2)}{2}.\nonumber
\eea
Inserting these back into (\ref{p2}), one finds
\bea
\asb P^2\as&=&\frac{1}{4}\ppink{2}\left|\Delta^{001}_{k_1 k_2}\right|^2\frac{\ok2-\ok1}{\ok1}\label{p2fin}\\
&&+\frac{1}{2}\ppin{k}\left|\Delta^{001}_{k B}\right|^2 \left(\sigma^2\ok{}+1+\frac{1}{4\sigma^2\ok{}}+\frac{Q_0\alpha^2}{\ok{}}\right)\nonumber\\
&=&\frac{1}{8}\ppink{2}\left|\Delta^{001}_{k_1 k_2}\right|^2\frac{\left(\ok2-\ok1\right)^2}{\ok1\ok2}\nonumber\\
&&+\frac{1}{2}\ppin{k}\left|\Delta^{001}_{k B}\right|^2 \left[\left(\sigma\sqrt{\ok{}}+\frac{1}{2\sigma\sqrt{\ok{}}}\right)^2+\frac{Q_0\alpha^2}{\ok{}}\right].\nonumber
\eea

The symbol $\Delta$ is independent of $g$ and $\sigma$.  Therefore the first term is of order $m^2$.  This means that this term, like the kink center of mass, yields a contribution to the momentum smearing of order the meson mass $m$.    But are these integrals finite?  For a gapped model, $\g_B(x)$ falls to zero exponentially, and so the integrals with $\Delta^{001}_{k B}$ converge.  In general $\Delta^{001}_{k_1 k_2}$ contains a $(k_1-k_2)\delta(k_1+k_2)$ term arising from the high $|x|$ tail of $\g_k(x)$, where it becomes a plane wave.  The $\delta$ function in each $\Delta$ is canceled by a factor of $\ok2-\ok1$ in (\ref{p2fin}).  In the $\phi^4$ \cite{mephi4} and Sine-Gordon models \cite{me2loop},  $\Delta^{001}_{k_1 k_2}$ also contains a term of the form $(k_2-k_1)^2 \csch(\pi(k_1+k_2)/m)/(\ok1\ok2)$.  The second order pole at $k_1=-k_2$ is removed by the second order zero in $(\ok2-\ok1)^2$ in (\ref{p2fin}).  $\Delta^2$ falls exponentially as $|k_1+k_2|$ increases, and so any divergence must occur along the strip at finite $k_1+k_2$ as $|k_1|$ goes to $\infty$.  However here there are four powers of $\omega_k$ in the denominator, and also $\ok2-\ok1$ shrinks, and so this contribution to the integral is also quite convergent.  Thus we conclude that, at least in the Sine-Gordon and $\phi^4$ models, these integrals are  convergent and so can, up to a constant of order unity, be estimated by the corresponding power of $m$ obtained from dimensional analysis.

What about the last line of (\ref{p2fin})?  As $\sigma$ has dimensions of mass${}^{-1/2}$, the terms are of order $m^3\sigma^2$, $m/\sigma^2$ and $m^2$ .    The bound (\ref{siglim}) implies that the first is less than $mQ_0\sim m^2/g^2$ while the second is greater than $m^2g^2$.   Thus the standard deviation of the momentum is bounded from below by $mg$ for wave packets of the form (\ref{sig}).

The total variance is
\bea
\asb\df^\dag P^2 \df\as-Q_0^2\alpha^2&=&\asb \left(P-\sqrt{Q_0}\pi_0\right)^2\as-Q_0^2\alpha^2\label{varfin}\\
&=&\frac{Q_0}{4\sigma^2}+\frac{1}{8}\ppink{2}\left|\Delta^{001}_{k_1 k_2}\right|^2\frac{\left(\ok2-\ok1\right)^2}{\ok1\ok2}\nonumber\\
&&+\frac{1}{2}\ppin{k}\left|\Delta^{001}_{k B}\right|^2 \left[\left(\sigma\sqrt{\ok{}}+\frac{1}{2\sigma\sqrt{\ok{}}}\right)^2+\frac{Q_0\alpha^2}{\ok{}}\right].\nonumber\\
&\sim&O\left(\frac{m}{g^2\sigma^2}\right)+O\left(m^2\right)+O\left({m^3}\sigma^2\right)+O\left(\frac{m}{\sigma^2}\right).\nonumber
\eea
The $O(m^2)$ term never dominates and, as $g<<1$, there is no range of parameters for which the $O(m/\sigma^2)$ term dominates.  The minimum of the variance is $O(m^2/g)$ which occurs when $\sigma\sim 1/\sqrt{mg}$ corresponding to $\sigma_{x_0}\sim \sqrt{g}/m$.  This corresponds to a spatial smearing which is smaller than the classical solution by of order $\sqrt{g}$.  It is just at the edge of the regime of validity (\ref{pertlim}) of our perturbative expansion in $g\phi_0^2$, but well within the semiclassical regime (\ref{siglim}).

\subsubsection{When is the Smearing Less Than the Momentum?}

This limits the kink rapidities to which our wave packets may be applied.  Clearly the rapidity must be much less than unity for the nonrelativistic approximation, which is implied by the semiclassical expansion, to apply.  However in the nonrelativistic regime the momentum is $Q_0\alpha\sim m\alpha/g^2$.  The condition $Q_0\alpha>>m/\sqrt{g}$, that the momentum exceeds the momentum spread, then yields
\beq
1>>\alpha>>g^{3/2}.
\eeq
Had this interval been empty, our choice of wave packet $\as$ would have needed to be revisited.  In particular, the momentum and kinetic energy satisfy
\beq
\frac{m}{g^2}>>Q_0\alpha>>\frac{m}{\sqrt{g}}\hsp
\frac{m}{g^2}>>Q_0\frac{\alpha^2}{2}>>mg.
\eeq
Note that this lower bound on the energy from smearing is smaller than the one-loop contribution to the energy $Q_1$, which is of order $m$, but it is larger than the two-loop contribution $mg^2$.  Thus, for a wave packet of the form (\ref{varfin}), it is not useful to consider two-loop corrections to energies, as these are subdominant to the smearing caused by the wave packet.

For smaller rapidities, the momentum width will exceed its central value for any semiclassical kink wave packet.   Note that there is no such lower bound on $\alpha$ using the nonnormalizable construction of Sec.~\ref{boostsez}, where semiclassical expansion converges, in the usual sense, to momentum eigenstates.

It is plausible that if we improved the wave packet $\as$ definition in (\ref{sig}), for example by using a higher order approximation to $|\alpha\rangle$ than $|\alpha\rangle_0$, the $\langle P^2\rangle$ term in (\ref{varfin}) would not be present or would be smaller.  This may allow us to extend the wave packet approach down to lower rapidities, nearing the bound of $\alpha\sim g^2$ from (\ref{lower}) where the kink momentum is of order the meson mass.  In this case the contribution of the wave packet smearing to the energy would be of the same order $mg^2$ as the two-loop corrections.

\section{Remarks}

Linearized soliton perturbation theory allows for fast and reliable calculations of quantities in soliton sectors of quantum field theories.  The limitation is that it is obtained via a linear expansion about a single base point in moduli space.   A Hamiltonian eigenstate is a superposition of solitons over the entire moduli space, and so this state necessarily extends beyond the validity of the expansion.  As a result, the applications of this method have been limited to expansions of states near the base point and quantities, like the energy spectrum, that are uniquely determined by the solution in any small region.

In this paper we extended linearized soliton perturbation theory to soliton states with momentum.   We did this both for Hamiltonian eigenstates, which are spread over the entire moduli space, and also for localized wave packets.  Our wave packets are normalizable, which means that, for a sufficiently small size, the linearized perturbation theory converges in the sense of an asymptotic series.  Furthermore, for the first time it allows us to compute matrix elements.

Now that we have both finite momentum and also normalizable states, our next task will be to compute form factors.  These will be unrelated to the form factors that are well known in the Sine-Gordon model \cite{weisz77,smirnov92,bab01}, which apply to Hamiltonian eigenstates.  Instead they will be form factors for solitons whose smearing is smaller than their classical size, which is arguably a more common situation in Nature than infinitely-extended Hamiltonian eigenstates.  It will, to our knowledge, be the first time that soliton form factors have been calculated in this strongly semiclassical regime.  

Beyond form factors, this formalism allows for a fast calculation of various matrix elements of interest.  For example, by including a $B^\dag$ on one side of a form factor, one arrives at a matrix element for the excitation of a normal mode during meson-kink scattering.  One can similarly calculate all of the matrix elements necessary to describe a number of aspects of meson-kink scattering, kink excitation, kink de-excitation or even the effects of quantum quenches on kinks.   However, the intrinsic smearing of our wave packets (\ref{sig}) implies that we will only be able to treat the scattering of nonrelativistic kinks with ultrarelativistic mesons.  In contrast, progress towards form factors of relativistic kinks has recently appeared in Ref.~\cite{andyff2}.

Another application is the construction of an effective moduli space Hamiltonians in models without Poincar\'e invariance, such as kinks in backgrounds with impurities \cite{muri}.  These depend on both the position and also the velocity in moduli space, and so can be derived by calculating the energies of moving kinks.  

The  extension of linearized soliton perturbation theory to states with momentum is a necessary step on the road to a treatment of explicitly time-dependent solitons.  A first quantum treatment of such solutions has recently been presented in Ref.~\cite{kovtun}.  Similarly, one could attempt to apply this formalism to theories with noncanonical kinetic terms.  Here the form of $H\p_2$ may differ.  Quantum corrections to kinks in such theories have recently been considered in Ref.~\cite{yuan21} with normal modes systematically investigated in Refs.~\cite{yuannor1,yuannor2}.

\appendix

\section{Delta Symbols}

We will introduce some notation
\beq
\Delta^{lmn}_{ij}=\int dx x^l \partial^m_x \g_i(x) \partial^n_x \g_j(x). \label{deldef}
\eeq
Not all of these are independent.  For example, integrating by parts 
\beq
\Delta^{001}_{ij}=-\Delta^{001}_{ji}
\eeq
and one easily sees that all $\Delta^{lmm}$ are symmetric, and that the symbol is symmetric under the interchange of $\{m,i\}$ with $\{n,j\}$.  Using the wave equation (\ref{sl}) one can show
\beq
\partial_x\left(\g_i(x)\partial_x \g_j(x)-\g_j(x)\partial_x \g_i(x)\right)=\g_i(x)\partial^2_x \g_j(x)-\g_j(x)\partial^2_x \g_i(x)=(\omega_i^2-\omega_j^2)\g_i(x)\g_j(x)
\eeq
and so, integrating by parts
\beq
\Delta^{100}_{ij}=\int dx x \g_i(x) \partial_x \g_j(x)=-\int dx \frac{\left(g_i(x)\partial_x \g_j(x)-g_j(x)\partial_x \g_i(x)\right)}{(\omega_i^2-\omega_j^2)}=\frac{2\Delta^{001}_{ij}}{(\omega_j^2-\omega_i^2)}. \label{did}
\eeq

Using the completeness (\ref{comp}) of the normal modes, one can prove a number of identities for bilinears of $\Delta$ symbols such as
\bea
\ppin{k\p}\Delta^{100}_{Bk\p}\Delta^{001}_{B -k\p}&=&\frac{1}{2}\\
\Delta^{100}_{BB}\Delta^{001}_{Bk}+\ppin{k\p}\left(\Delta^{100}_{Bk\p}\Delta^{001}_{-k\p k}+\Delta^{100}_{k\p k}\Delta^{001}_{-k\p B}\right)&=&0\nonumber\\
\Delta^{100}_{B(k_1}\Delta^{001}_{k_2) B}+\ppin{k\p}\Delta^{100}_{(k_1 k\p}\Delta^{001}_{k_2) -k\p}&=&\pi\delta(k_1+k_2)\nonumber
\eea
where we remind the reader that $\dint$ includes a sum over all shape modes and the parenthesis on indices represent symmetrization with a factor of $1/2$.   Similarly one can show
\bea
\ppin{k\p}\Delta^{111}_{Bk\p}\Delta^{001}_{B -k\p}&=&\frac{\Delta^{011}_{BB}}{2}\\
\Delta^{111}_{BB}\Delta^{001}_{Bk}+\ppin{k\p}\left(\Delta^{111}_{Bk\p}\Delta^{001}_{-k\p k}+\Delta^{111}_{k\p k}\Delta^{001}_{-k\p B}\right)&=&\Delta^{011}_{Bk}\nonumber\\
\Delta^{111}_{B(k_1}\Delta^{001}_{k_2) B}+\ppin{k\p}\Delta^{111}_{(k_1 k\p}\Delta^{001}_{k_2) -k\p}&=&\frac{\Delta^{011}_{k_1k_2}}{2}.\nonumber
\eea

\section* {Acknowledgement}

\noindent
JE is supported by the CAS Key Research Program of Frontier Sciences grant QYZDY-SSW-SLH006 and the NSFC MianShang grants 11875296 and 11675223.   JE also thanks the Recruitment Program of High-end Foreign Experts for support.

\end{document}